\documentclass[amsmath,amssymb,12pt]{iopart}
\usepackage{times}
\usepackage{graphicx}
\usepackage{dcolumn}
\usepackage{bm}
\usepackage{amsfonts}
\usepackage{color}

\newcommand{\F}[1]{\ensuremath{\mathbf{#1}}}
\newcommand{\sign}{\mathrm{sgn}}

\newcommand{\op}{\hat}
\newcommand{\Ci}{\mathrm{i}} 	

\newcommand{\eqref}[1]{\ref{#1}}
\newcommand{\pdiff}[1]{\partial_{#1}}

\newcommand{\opHoneB}{{\op H}_{\mathrm{1B}}}
\newcommand{\opUnit}{\mathbf{1}}
\newcommand{\text}[1]{\mathrm{#1}}
\newcommand{\Trace}{\mathrm{Tr}}
\newcommand{\const}{\text{const.}}
\newcommand{\fdiffer}[2]{\frac{\delta #1}{\delta #2}}
\newcommand{\fdiff}{\delta}
\newcommand{\trace}{\mathrm{tr}}
\newcommand{\diff}{\mathrm{d}}
\newcommand{\pDiff}[1]{\mathcal D_{#1}}
\newcommand{\toward}{\rightarrow}
\newcommand{\pdiffer}[2]{\frac{\partial #1}{\partial #2}}
\newcommand{\laplace}{\vec\nabla^2}
\newcommand{\unit}[1]{\mathrm{#1}}
\def\s0#1#2{\mbox{\small{$ \frac{#1}{#2} $}}}
\def\0#1#2{\frac{#1}{#2}}
\def\eq#1{(\ref{#1})}

\begin{document}

\title[2PI nonequilibrium versus transport equations]{2PI nonequilibrium versus transport equations for an ultracold Bose gas}
\author{Alexander Bransch\"adel and Thomas Gasenzer}
\address{Institut f\"ur Theoretische Physik, Universit\"at Heidelberg,\newline Philosophenweg 16, 69120 Heidelberg, Germany}
\date{\today}

\begin{abstract} 
\noindent
The far-from-equilibrium dynamics of an ultracold, one-dimensional Bose gas is studied. 
The focus is set on the comparison between the solutions of fully dynamical evolution equations derived from the 2PI effective action and their corresponding kinetic approximation in the form of Boltzmann-type transport equations. 
It is shown that during the time evolution of the gas a kinetic description which includes non-Markovian memory effects in a gradient expansion becomes valid.
The time scale at which this occurs is shown to exceed significantly the time scale at which the system's evolution enters a near-equilibrium drift period where a fluctuation dissipation relation is found to hold and which would seem to be predestined for the kinetic approximation.

\end{abstract}
\pacs{03.70.+k, 03.75.Kk, 05.20.Dd, 05.70.Ln, 11.15.Pg, 51.10.+y\newline\newline
HD--THEP--08--02
\newline
\vfill
\noindent
Email: {\tt A.Branschaedel@thphys.uni-heidelberg.de},\newline 
\phantom{Email:} {\tt T.Gasenzer@thphys.uni-heidelberg.de}\newline
}
\maketitle
\section{Introduction}
\label{sec:intro}

The dynamical evolution of ultracold atomic quantum gases driven far out of equilibrium is a subject of intensively growing interest.
Precision measurement techniques for many-body observables have been and are being developed with vigourous effort.
This technology has triggered a demand for progress in theoretically describing non-equilibrium quantum many-body dynamics of strongly interacting systems beyond mean-field approximations.
Recent highlights of this development include, e.g., the variation and enhancement of the atomic interactions on the basis of Feshbach scattering resonances \cite{Inouye1998a},
the achievement of strongly correlated regimes within optical lattices \cite{Greiner2002a},
as well as the quench dynamics of spinor Bose-Einstein condensates \cite{Sadler2006a}.

  To obtain the full dynamics of a quantum many-body system, it is convenient to formulate an initial-value problem that describes the time evolution by means of in general coupled equations of motion for time dependent correlation functions and specific values for these functions at initial time.
Alternatively, the time-evolution of many-body systems is very often described in terms of kinetic or transport equations \cite{KadanoffBaym1962a}.  
In general, the aim of quantum kinetic theory is to find evolution equations for distribution functions $f(\F x, t)$ or $\tilde f(\F p,t)$, interpreted, e.g., as particle number densities in $\F x$- or $\F p$-space. 
These distribution functions can then be used to derive various transport properties of the many-particle system as, e.g., current of charge or energy. For that reason, the evolution equations of $f$ are generally referred to as transport equations.  

Kinetic descriptions usually neglect the effect of correlations between different times of the evolution, i.e., they build, to a certain extent, on a Markovian approximation. 
In particular, they neglect the initial dynamics directly after a change in the boundary conditions which drive the system out of equilibrium.
This shortcoming is cured in dynamical approaches in which coupled equations of motion for the correlation functions are derived to describe the time evolution starting from a specific initial state.
The built-up of correlations beyond the kinetic approximation, in these equations, is taken into account by means of non-Markovian integrations over the evolution history of the correlation functions.
   
The focus of this work is set on the question to what extent transport theory can be used to quantitatively describe the dynamics of an ultracold Bose gas at different stages of its evolution from the initial to the equilibrium state. 
Determining the range of validity of transport equations by comparing to the full quantum dynamic evolution helps forming a solid foundation for the application of transport equations which, for practical problems, often is technically less demanding.
   
In the present work we use the two-particle irreducible (2PI) effective action \cite{Luttinger1960a}
which is derived from a functional integral representation of the quantum many-body system. 
From this action, self-consistent quantum-dynamic equations for the mean field as well as the two-point correlation functions are obtained \cite{Berges:2001fi,Aarts2002b}.
These have been applied successfully for the description of far-from-equilibrium dynamics and thermalization \cite{Berges:2000ur,Berges:2001fi,Cooper:2002qd,Berges:2002wr,Juchem:2003bi,Arrizabalaga:2005tf,Gasenzer:2005ze,Berges2007a}.
For our investigations, we numerically compare the time evolution determined with dynamic equations derived from the 2PI effective action with the corresponding transport equations. 
These transport equations represent an approximation to the full dynamic equations.
The derivation of transport equations has been discussed extensively in the literature. 
In the context of non-relativistic systems, in particular cold atomic gases cf., e.g., Refs.~\cite{KadanoffBaym1962a,Proukakis1996a,Shi1998a,Gardiner1997a,Giorgini1997a,Proukakis1998a,Walser1999a,Gardiner2000a,Walser2000a,Rey2005a,Baier2005a}. 
See also Refs.~\cite{Danielewicz1984a,Mrowczynski:1992hq,
Calzetta1988a,Blaizot:2001nr,Berges:2002wt,Prokopec:2003pj,Lindner:2005kv,
Berges:2005md} in the context of relativistic physics.
A comparison of dynamic equations with their kinetic approximation for relativistic dynamics has been given in Refs.~\cite{Danielewicz1984a,KohlerHS1995a,Morawetz1999a,Juchem:2003bi,Lindner:2005kv,Berges:2005md}. 
Here, we derive transport equations in leading order (LO) and next-to-leading order (NLO) of a gradient expansion. 
This is essentially achieved in four steps: 
(a) the previously obtained 2PI dynamic equations for the two-point correlation functions which are expressed with respect to absolute coordinates are rewritten with respect to center and relative coordinates. 
This allows (b) for a gradient expansion with respect to center coordinates. 
(c) Furthermore, knowledge regarding the details of the initial state is neglected. 
(d) These steps allow for a Wigner transformation, i.e., a Fourier transformation with respect to relative coordinates, which yields the desired transport equations. 
We then use the results obtained from solving the 2PI dynamic equations to study the range of validity of the transport equations.
  
While the solution of an initial-value problem requires to formulate a dynamical instead of a kinetic theory, a further important aspect is to properly take into account higher order correlations in the description of strongly correlated systems.
Far from equilibrium, perturbative mean-field approximations such as in Gross-Pitaevskii \cite{Gross1961a}
and Hartree-Fock-Bogoliubov \cite{Hartree1928a}
theory cease to be valid descriptions.  
The dynamics of a self interacting ultracold gas is well-described by such mean-field approximations only as long as fluctuations and the interaction strength are small such that expansions in a small coupling parameter are justified.
This is generally fulfilled for weakly interacting systems close to equilibrium.
For strongly fluctuating systems this approach, however, becomes insufficient when requiring summations of infinite series of perturbative processes. 
But even close to equilibrium, fluctuations can play an important role on large time scales.

In this article we consider a non-perturbative approximation which reaches substantially beyond the Hartree-Fock-Bogoliubov mean field theory.
This non-perturbative approach is based on a systematic expansion of the 2PI effective action in powers of the inverse number of field components ${\cal N}$ \cite{Berges:2001fi,Aarts2002b}. 
It has recently been used to describe the dynamics of an ultracold atomic Bose gas far from equilibrium  \cite{Gasenzer:2005ze,Berges2007a} and its non-perturbative character has been reconfirmed in the framework of a functional renormalization group approach \cite{Gasenzer:2007za}.
The 2PI $1/{\cal N}$ expansion to next-to-leading order yields dynamic equations which contain direct scattering, memory and ``off-shell'' effects. 
It allows to describe far-from-equilibrium dynamics as well as the late-time approach to quantum thermal equilibrium.
Recently, these methods have allowed important progress in describing the dynamics of strongly interacting relativistic systems far from thermal equilibrium for bosonic \cite{Berges:2001fi,Berges:2002cz,Mihaila2001a,Cooper:2002qd,Arrizabalaga:2004iw,Aarts:2006cv} as well as fermionic degrees of freedom \cite{Berges:2002wr,Berges:2004ce}. 

Our article is organized as follows:
In Section \ref{sec:DynEq} we recall the functional description of the quantum many-body dynamics on the basis of the 2PI effective action and present the set of coupled dynamic equations for the two-point correlation functions.
In Section \ref{sec:TransportEq} we derive, by transforming to Wigner coordinates and performing a gradient expansion, the set of transport equations which we compare, in Section \ref{sec:DynvsTrans}, numerically with the 2PI dynamic equations.
Our conclusions are drawn in Section \ref{sec:Concl}.

\section{Dynamical evolution equations}
\label{sec:DynEq}

The evolution of single-particle observables such as the spatial density and momentum distributions are most conveniently described through the spectral and statistical two-point functions, denoted as $\rho$ and $F$, respectively \cite{Aarts:2001qa}:
  \begin{eqnarray}
    F_{ab}(x,y) 
      &=& \frac 1 2 \langle\lbrace\Phi_a(x),\Phi_b(y)
                  \rbrace\rangle_c 
      = \frac 1 2 \langle\lbrace\Phi_a(x),\Phi_b(y)
                  \rbrace\rangle - \phi_a(x)\phi_b(y),~~ \\
    \rho_{ab}(x,y) 
      &=& \Ci \langle[\Phi_a(x),\Phi_b(y)]\rangle_c
      = \Ci \langle[\Phi_a(x),\Phi_b(y)]\rangle.
  \end{eqnarray}
Here, $\langle\dots\rangle$ implies a trace over the density matrix describing the full quantum state at the initial time, and $\langle\dots\rangle_c$ is a shorthand for the corresponding connected correlation function or cumulant.
As defined in more detail below, we will consider a non-relativistic theory for a complex scalar field with  ${\cal N}=2$ components describing an ultracold Bose gas, and we assume pointlike interactions between the atoms.
Hence, all fields and $n$-point functions acquire indices $a,b,...\in\{1,2\}$.

From the above definition it is clear that the full propagator function $G$, defined as the expectation value of the time-ordered product of field operators, can be written in terms of the statistical and spectral functions as follows:
  \begin{eqnarray}
\label{eq:DE:decompositionG}
    G_{ab}(x,y) 
    &=& \langle{\cal T_{\cal C}}\Phi_a(x),\Phi_b(y)
                  \rangle_c
		  \nonumber\\
    &=& F(x,y)-\frac \Ci 2 \rho(x,y)\,\mathrm{sgn}_{\mathcal C}(x_0-y_0).
  \end{eqnarray}
Our aim is to solve an initial-value problem for the two-point function, such that we need to work with the Schwinger-Keldysh closed time path (CTP) $\cal C$ \cite{Schwinger1961a}
from the initial time along the positive time axis to the largest time appearing in the product of field operators, and back to the initial time.
$\cal T_{\cal C}$ denotes time ordering along $\cal C$, and the sign function $\mathrm{sgn}_{\cal C}$ evaluates to $1$ ($-1$) for $x_0$ later (earlier) than $y_0$ along the CTP.

While the spectral function $\rho$ encodes the excitation spectrum of the theory, the statistical propagator $F$ contains information about the occupation numbers.
Roughly speaking, $\rho$ makes explicit what states are available and $F$ how often they are occupied.
Far away from equilibrium, $F$ and $\rho$ are in general independent functions, while close to thermal equilibrium, they are connected by a fluctuation dissipation relation \cite{KadanoffBaym1962a,Aarts:1997kp}.

For $x_0=y_0$ the equal time commutation relations for the field operators $\Phi$ imply
  \begin{equation}
    \rho(x,y)\big\vert_{x_0=y_0} = -\Ci \sigma_2 \delta(\F x-\F y).
\label{eq:DE:rhoEqualTime}
  \end{equation} 

The exact dynamical evolution equations for the statistical and spectral functions can be derived from the 2PI effective action \cite{Berges:2001fi,Aarts2002b} and read, for a vanishing field expectation value $\phi_a(x)=\langle\Phi_a(x)\rangle\equiv0$:
  \begin{eqnarray}
\label{eq:TE:motionF1}
    \pdiff{x_0}F(x,y) 
    &=&\, \Ci\sigma_2\Bigg\lbrace\Big[\opHoneB(x)+\Sigma^{(0)}(x)\Big]F(x,y) 
    \nonumber\\
    &&+\int_{z,0}^{x_0}\Sigma^\rho(x,z)F(z,y)-\int_{z,0}^{y_0} \Sigma^F(x,z)\rho(z,y)\Bigg\rbrace, 
   \\
\label{eq:TE:motionRho1}
    \pdiff{x_0}\rho(x,y) 
    &=&\, \Ci\sigma_2\Bigg\lbrace\Big[\opHoneB(x)+\Sigma^{(0)}(x)\Big] \rho(x,y) 
         +\int_{z,y_0}^{x_0}\Sigma^\rho(x,z)\rho(z,y)\Bigg\rbrace.
  \end{eqnarray} 
Here, $\sigma_2$ is the Pauli 2-matrix, and we have suppressed the explicit notion of field indices, i.e., all functions are understood to be $2\times2$-matrices.
We use the notation $\int_{z,y_0}^{x_0}\equiv\int_{y_0}^{x_0}\diff z_0\int\diff^D z$.
Equations \eq{eq:TE:motionF1} and \eq{eq:TE:motionRho1} represent an initial value problem, i.e., they give the time evolution of the functions $F$ and $\rho$ for given initial values $F(t_0,\mathbf{x};t_0,\mathbf{y})$ and $\rho(t_0,\mathbf{x};t_0,\mathbf{y})$.
The memory integrals on the right hand sides take into account the build up of higher-order correlations during the evolution and ensure the equations to respect causality.
Since the equations are exact, they are equivalent to any other representation such as dynamical Kadanoff-Baym or Schwinger-Dyson equations for the time-dependent two-point functions.

The spectral ($\Sigma^\rho$) and statistical ($\Sigma^F$) parts of the self-energy, as well as the local energy shift $\Sigma^{(0)}$ are obtained from the proper self-energy $\Sigma$ which sums all one-particle irreducible (1PI) diagrams, in a similar way as in Eq.~(\ref{eq:DE:decompositionG}):
  \begin{eqnarray}
  \label{eq:SigmaDecomp}
    \Sigma(x,y) 
    &=& -\Ci\Sigma^{(0)}(x)\delta(x-y) +\bar\Sigma(x,y), \\
  \label{eq:SigmabarDecomp}
    \bar\Sigma(x,y) 
    &=& \Sigma^F(x,y)-\frac \Ci 2 \Sigma^\rho(x,y)\,\sign_{\mathcal C}(x_0-y_0).
  \end{eqnarray} 

\begin{figure}[tb]
\begin{center}
    \includegraphics[width=0.67\textwidth]{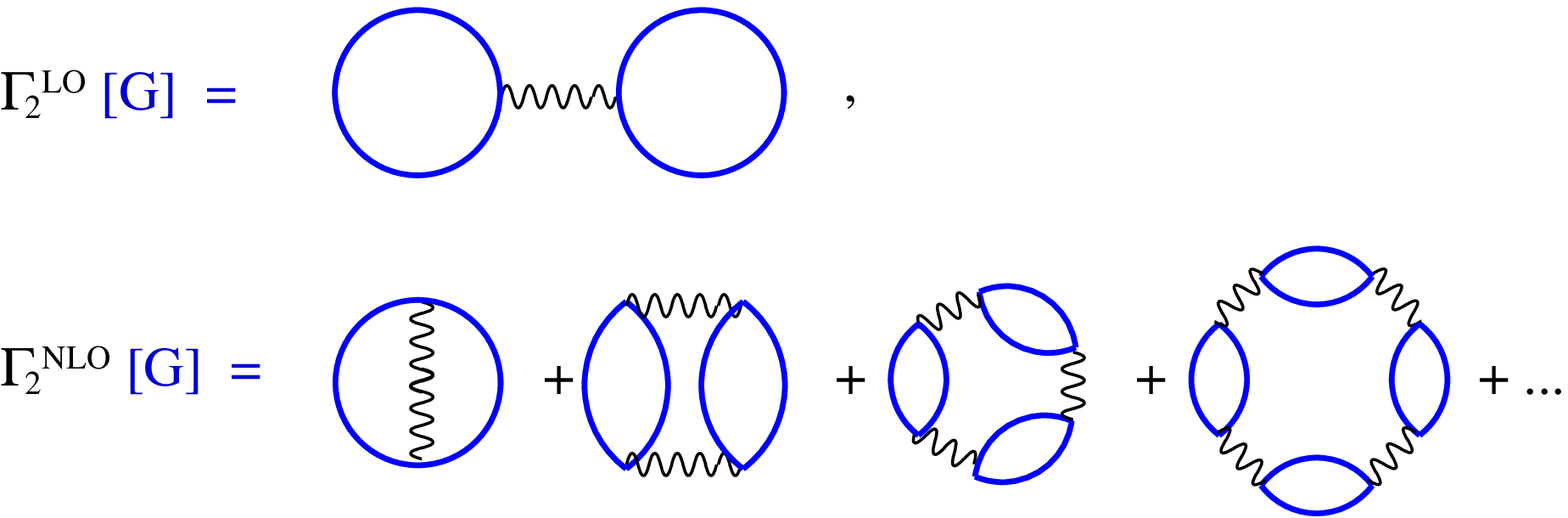}
    \caption{
(Color online) The diagrams contributing, in $\mathrm{LO}$ and $\mathrm{NLO}$ of the $1/\mathcal N$-expansion, to the $\mathrm{2PI}$ part $\Gamma_2[\phi, G]$ of the $\mathrm{2PI}$ effective action. At each vertex, it is summed over double field indices and integrated over double space-time variables. The wiggly lines represent the bare potential, the blue solid lines the full propagator $G$.}
\label{fig:BC:1overNGraphs}
\end{center}
\end{figure}

To close the set of dynamic equations (\ref{eq:TE:motionF1}) and (\ref{eq:TE:motionRho1}), the self energy has to be specified.
We derive the self energy from the beyond-one-loop contribution $\Gamma_2$ to the 2PI effective action $\Gamma$,
  \begin{equation}
    \Gamma[\phi,G] = \Gamma^{\text{one-loop}}[\phi, G]
                    +\Gamma_2[\phi,G],
\label{eq:BC:GammaFull}
  \end{equation} 
  \begin{equation}
\label{eq:BC:Gamma1Loop}
    \Gamma^{\text{one-loop}}[\phi, G]
       = S[\phi] 
       + \frac\Ci2\Trace\big\lbrace\ln(G^{-1})\big\rbrace
       + \frac\Ci 2\Trace\big\lbrace G_0^{-1}[\phi]G\big\rbrace
       + \const,
  \end{equation} 
by taking the functional derivative with respect to the two-point function $G$,
  \begin{equation}
    \Sigma_{ab}(x,y) = 2\Ci\fdiffer{\Gamma_2[\phi,G]}{G_{ab}(x,y)}.
\label{eq:BC:Selfenergy}
  \end{equation} 
In practice, deriving $\Sigma$ is not possible without making approximations since $\Gamma_2$ consists of an infinite number of 2PI diagrams constructed from the bare vertex as well as the full 2-point function $G$.
In Eq.~(\ref{eq:BC:Gamma1Loop}), $G_0^{-1}$ is the inverse classical propagator obtained from the classical action $S[\phi]$, which defines the model to be considered, as
  \begin{eqnarray}
    \Ci G_{0,ab}^{-1}(x,y) 
      &=& \frac{\fdiff^2S[\phi]}
             {\fdiff\phi_a(x)\fdiff\phi_b(y)}.
\label{eq:BC:classicalPropagator}
  \end{eqnarray} 
In the following we choose the $1/\cal N$ expansion to next-to-leading order (NLO) where $\Gamma_2$ is given as
  \begin{equation}
  \label{eq:Gamma2NLO}
    \Gamma_2[\phi,G]=\Gamma_2^{\mathrm{LO}}[G]+\Gamma_2^{\mathrm{NLO}}[\phi, G].
  \end{equation} 
The leading order (LO) contribution to $\Gamma_2$ consists of a $\phi$-independent two-loop graph \cite{Gasenzer:2005ze}
  \begin{equation}
    \Gamma_2^{\mathrm{LO}}[G] = -\frac g{4\mathcal N}\int_x \trace \lbrace G(x,x)\rbrace\trace\lbrace G(x,x)\rbrace
\label{eq:BC:Gamma2LO}
  \end{equation} 
which is shown in Figure \ref{fig:BC:1overNGraphs}(a). 
The NLO contributions, Fig.~\ref{fig:BC:1overNGraphs}(b), can be summed up analytically, yielding 
  \begin{equation}
    \Gamma_2^{\mathrm{NLO}}[G] = \frac\Ci 2\Trace\lbrace\ln B[G]\rbrace, 
\label{eq:BC:Gamma2NLO}
  \end{equation} 
  with 
  \begin{equation}
    B(x,y;G) =\delta(x-y)+\frac{\Ci g}{\mathcal N}\trace\lbrace G(x,y)G(y,x)\rbrace
\label{eq:BC:Gamma2B}.
  \end{equation} 
%
Inserting Eqs.~(\ref{eq:Gamma2NLO})--(\ref{eq:BC:Gamma2B}) into (\ref{eq:BC:Selfenergy}) and using Eqs.~(\ref{eq:SigmaDecomp}) and (\ref{eq:SigmabarDecomp}) one obtains
  \begin{eqnarray}
    \Sigma^{(0)}(x) 
      &=& g \big[F(x,x)+\frac 1 2 \trace\lbrace F(x,x)\rbrace\big],
\label{eq:DE:Sigma0Def}
    \\
    \Sigma^{F}(x,y)
      &=& -g \Big\lbrace I^F(x,y)F(x,y)
          -\frac {1}{4} I^\rho(x,y) \rho(x,y)
    \Big\rbrace,
  \label{eq:DE:SigmaF}
    \\
    \Sigma^{\rho}(x,y)
      &=& -g \Big\lbrace I^\rho(x,y)F(x,y) 
          + I^F(x,y)\rho(x,y)
     \Big\rbrace.
  \label{eq:DE:SigmaRho}
  \end{eqnarray} 
The functions $I^{F,\rho}$ are obtained in analogy to Eq.~(\ref{eq:SigmabarDecomp}) from the function
  \begin{eqnarray}
    I(x,y) 
     &=& \frac g{\mathcal N} \left[\trace\lbrace G(x,y) G(y,x)\rbrace 
     -\Ci
     \int_z I(x,z;G)\trace\lbrace G(z,y)G(y,z)\rbrace\right]
\label{eq:BC:Gamma2I}
  \end{eqnarray} 
which results from taking the derivative of $\Gamma^\mathrm{NLO}$ with respect to $G$ and are given as
  \begin{eqnarray}
    I^F(x,y) 
      &=& \frac{g}{\mathcal N}\Bigg[F(x,y)^2-\frac 1 4\rho(x,y)^2
          - \int_{z,0}^{x_0} I^\rho(x,z)\left( F(z,y)^2-\frac 1 4 \rho(z,y)^2\right)
      \nonumber \\
      && +2\int_{z,0}^{y_0}I^F(x,z) F_{ab}(z,y)\rho_{ab}(z,y)
         \Bigg],
\label{eq:DE:IF}
      \\
    I^\rho(x,y) 
      &=& \frac{2g}{\mathcal N}\left[ F_{ab}(x,y)\rho_{ab}(x,y)-\int_{z, y_0}^{x_0} I^\rho(x,z)F_{ab}(z,y)\rho_{ab}(z,y)\right]
\label{eq:DE:IRho}
  \end{eqnarray} 
where $F^2\equiv F_{ab}F_{ab}$ etc., and where a decomposition of $I$ similar to that in Eq.~(\ref{eq:DE:decompositionG}) has been applied.

This approximation has been shown to describe thermalization of a one-dimensional Bose gas starting far from thermal equilibrium \cite{Gasenzer:2005ze,Berges2007a} as well as for relativistic field theories \cite{Berges:2001fi,Berges:2002cz,Cooper:2002qd,Arrizabalaga:2004iw,Berges:2002wr,Berges:2004ce}.
It includes off-shell scattering processes which allow the system to exhibit damping, even in one spatial dimension.
We point out, that the dynamic equations sum over the full past time evolution of the correlation functions and therefore include memory.
Hence, no derivative expansion has been applied yet, which will constitute the main step to arrive at transport equations.

\section{Transport equations}
\label{sec:TransportEq}
In this section we derive the transport equations in leading order (LO) and next-to-leading order (NLO) of a gradient expansion. 
We rewrite the exact dynamic equations given in the last section in order to get evolution equations with respect to centre and relative coordinates, providing a starting point for a gradient expansion and transformation to Wigner space. 
For simplicity we consider a system with spatially homogeneous initial conditions for the two-point function:
\begin{equation}
  F_{ab}(0,\mathbf{x},0,\mathbf{y})
  \equiv  F_{0,ab}(\mathbf{x},\mathbf{y})
  = \delta_{ab}f_0(\mathbf{x}-\mathbf{y}).
  \label{eq:IniCondF}
\end{equation}
The initial values of the spectral function are fixed by the commutation relations, cf.~Eq.~(\ref{eq:DE:rhoEqualTime}):
\begin{equation}
  \rho_{0,ab}(\mathbf{x},\mathbf{y})
  = -i\sigma_{2,ab}\delta(\mathbf{x}-\mathbf{y}).
  \label{eq:IniCondRho}
\end{equation}

To meet the requirements of the Fourier transformed equations in Wigner space on the one hand and to benefit from the spatial homogeneity of the system on the other hand, we rewrite the equations in terms of relative and centre coordinates
  \begin{equation}
    X = \frac{x+y}2,~ s = x-y ~~ \Leftrightarrow ~~ x = X+s/2,~ y = X-s/2
  \end{equation} 
where the two-point functions are to be transformed as, e.g.,
      $F(x,y) \mapsto F(X,s)$, 
      $\Sigma^\rho(x,z) \mapsto \Sigma^\rho(X+s'/2,s-s')$, 
      $F(z,y) \mapsto F(X+(s'-s)/2, s')$.
  We additionally introduce
  \begin{equation}
    s' = z-y
  \end{equation} 
  which later on serves as the integration variable in the memory integrals.

Using these definitions in Eqs.~(\ref{eq:TE:motionF1}), (\ref{eq:TE:motionRho1}), and adding these equations to the corresponding equations for the time derivatives with respect to $y_0$, one obtains differential equations with respect to the centre time coordinate $X_0$:
  \begin{eqnarray}
   \pdiff{X_0}F(X,s) &=& \Ci\sigma_2 
         \Bigg\lbrace
           \Big[\opHoneB(X+\frac s 2)-\opHoneB(X-\frac s 2)
	   \nonumber \\ 
     && ~~~~~~~~~~ +\Sigma^{(0)}(X+\frac s 2) -\Sigma^{(0)}(X-\frac s 2)\Big]F(X,s) \nonumber \\
     && +\int_{s'}\theta(X_0+s'_0-\frac{s_0}2)\Big[\Sigma^R(X+\frac {s'} 2 , s-s') F(X-\frac{s-s'}2, s') \nonumber \\
     && ~~~~~~~~~~ - G^R(X+\frac{s'}2, s-s') \Sigma^F(X-\frac{s-s'}2,s') \nonumber \\
     && ~~~~~~~~~~ +\Sigma^F(X+\frac {s'} 2 , s-s')G^A(X-\frac{s-s'}2, s') \nonumber \\
     && ~~~~~~~~~~ - F(X+\frac{s'}2, s-s') \Sigma^A(X-\frac{s-s'}2,s')\Big]\Bigg\rbrace,
\label{eq:TE:FAbs}
  \end{eqnarray}
  \begin{eqnarray}
    \pdiff{X_0}\rho(X,s) 
     &=& \Ci\sigma_2 \Big\lbrace 
       \big[\opHoneB(X+\frac s 2)-\opHoneB(X-\frac s 2) 
     \nonumber \\
     && ~~~~~~~~~~ +\Sigma^{(0)}(X+\frac s 2) -\Sigma^{(0)}(X-\frac s 2) \big]\rho(X,s) \nonumber \\
     && +\int_{s'}\big[ \Sigma^R(X+\frac {s'}2,s-s') \rho(X-\frac{s-s'}2,s') \nonumber \\
     && ~~~~~~~~~~ -\rho(X+\frac{s'}2,s-s')\Sigma^A(X-\frac{s-s'}2,s') \nonumber \\
     && ~~~~~~~~~~ +\Sigma^\rho(X+\frac{s'}2,s-s')G^A(X-\frac{s-s'}2, s') \nonumber \\ 
     && ~~~~~~~~~~ -G^R(X+\frac{s'}2,s-s')\Sigma^\rho(X-\frac{s-s'}2, s')
     \big]
    \Big\rbrace.
\label{eq:TE:rhoAbs}
  \end{eqnarray} 
Note that, introducing retarded and advanced Greens functions and self-energies,
  \begin{eqnarray}
\label{eq:TE:rhoAR}
    G^R(x,y)
    &=&\theta(x_0-y_0)\rho(x,y), ~ G^A(x,y)=-\theta(y_0-x_0)\rho(x,y), \\
\label{eq:TE:SigmaAR}
    \Sigma^R(x,y)
    &=&\theta(x_0-y_0)\Sigma^\rho(x,y), ~ \Sigma^A(x,y)=-\theta(y_0-x_0)\Sigma^\rho(x,y), \end{eqnarray} 
allowed us to send the integration limits of $s'_0$ in Eqs.~(\ref{eq:TE:FAbs}), (\ref{eq:TE:rhoAbs}) to $\pm\infty$.
While the above equations have been obtained by adding the equations for $\partial_{x_0}F(x,y)$ and $\partial_{y_0}F(x,y)$, etc., a second set of equations for the derivatives of $F(X,s)$ and $\rho(X,s)$ with respect to $s_0$ results when subtracting the respective expressions.
These equations relate the two-time Greens functions to the single-time ones and are provided and discussed further in \ref{app:DEwrts0}.

\subsection{Approximations}
\label{sec:Approx}
With the aim to derive transport equations, we apply the following approximations:
\\

({\it i}) 
The $\theta$-function is neglected in the evolution equations for $F$, Eqs.~(\eqref{eq:TE:FAbs}), (\eqref{eq:TE:FRel}), taking into account that the correlations disappear for large relative times. 
This corresponds to sending the initial time $t_0$ to the infinite past. 
Note that, since an interacting system could have reached equilibrium at any finite time, transport equations are initialized by specifying $F$ and $\rho$ at a finite time using the equations with $t_0\to-\infty$ as approximate description \cite{Berges:2005md}. 
\\

({\it ii}) 
We apply a gradient expansion with respect to the centre coordinates $X$.
\\

\subsection{Gradient expansion}
\label{sec:GradExp}
In leading order of the gradient expansion, taking into account the homogeneous initial conditions, one obtains from Eqs.~(\ref{eq:TE:FAbs}) and (\ref{eq:TE:rhoAbs}) the evolution equations
  \begin{eqnarray}
\label{eq:TE:motionF_LO}
   \pdiff{X_0}F(X,s) 
    &=& \Ci\sigma_2 
          \int_{s'}\Big[F(X, s-s')\Sigma^\rho(X,s')-\rho(X,s-s')\Sigma^F(X, s')\Big],
    \\
    \pdiff{X_0}\rho(X,s) &=& 0.
  \end{eqnarray} 
Up to next-to-leading order in the gradient expansion these equations receive additional corrections as follows:
  \begin{eqnarray}
    \lefteqn{\pdiff{X_0}F(X,s)
       =\mathrm{LO}+\Ci\sigma_2
           \Bigg\lbrace 
              \big[s_0\pdiff{X_0} \Sigma^{(0)}(X)\big] F(X,s)} 
    \nonumber\\  
    && +\frac 1 2 \int_{s'}\Big [
        [s'_0\pdiff{X_0}\Sigma^+(X, s-s')]F(X,s')
        -\Sigma^+(X, s')[s'_0\pdiff{X_0} F(X,s-s')] \nonumber \\
   && ~~ -[s'_0\pdiff{X_0} \rho^+(X,s-s')]\Sigma^F(X,s') + \rho^+(X,s')[s'_0\pdiff{X_0} \Sigma^F(X,s-s')]\Big]\Bigg\rbrace,
   \nonumber \\
\label{eq:TE:motionF_NLO}
  \end{eqnarray} 
  \begin{eqnarray}
    \lefteqn{\pdiff{X_0}\rho(X,s)=\Ci\sigma_2\Bigg\lbrace \big[s_0\pdiff{X_0}\Sigma^{(0)}(X)\big] \rho(X,s)} \nonumber\\  
    &&+\frac 1 2 \int_{s'}\Big [[s'_0\pdiff{X_0}\Sigma^+(X, s-s')]\rho(X,s')
     -\Sigma^+(X, s')[s'_0\pdiff{X_0} \rho(X,s-s')] \nonumber \\
   && ~~ -[s'_0\pdiff{X_0} \rho^+(X,s-s')]\Sigma^\rho(X,s') + \rho^+(X,s')[s'_0\pdiff{X_0} \Sigma^\rho(X,s-s')]\Big]\Bigg\rbrace,
   \nonumber \\
  \end{eqnarray} 
   where  
  \begin{eqnarray}
    \rho^+(X,s) &=& G^R(X,s)+G^A(X,s),
  \label{eq:rhoplus}
    \\
    \Sigma^+(X,s) &=& \Sigma^R(X,s)+\Sigma^A(X,s)
  \label{eq:Sigmaplus}
  \end{eqnarray} 
and LO denotes the leading-order terms given by the right-hand side of Eq.~(\ref{eq:TE:motionF_LO}).

\subsection{Transformation to Wigner space}
\label{sec:TransWigner}
  The transformation to Wigner space involves a Fourier transformation with respect to the relative coordinate $s$
  \begin{equation}
    \tilde F(X,p) = \int_{s} \e^{\Ci ps} F(X,s), ~
    \tilde \Sigma^R(X,p) = \int_{s} \e^{\Ci ps} \Sigma^R(X,s),
  \end{equation} 
etc., where $ps=p_0 s_0-\F p\cdot\F s$.

For a spatially homogeneous system one finds that the diagonal matrix elements of $\tilde F$ are purely real, while the off-diagonal matrix elements are purely imaginary. 
Furthermore, we find $\tilde F^\mathrm{T}(X,p) = \tilde F^*(X,p)=\tilde F(X,-p)$. 
For $\tilde\rho$, the diagonal matrix elements are imaginary and the off-diagonal elements real. 
As compared with $\tilde F$, the function $\tilde\rho$ changes sign under the transposition, $\tilde\rho(X,p)^\mathrm{T} = -\tilde \rho^*(X,p)=-\tilde \rho(X,-p)$.

  We apply the transformation to the previously derived equations of motion and have to take into account the integration limits when interchanging the derivative with respect to the centre time with the relative time integration. 
 We thus introduce the derivative operator
   \begin{eqnarray}
    \pDiff{X_0}[\tilde F(X,p)] 
    &=& 
     \int_{-2X_0}^{2X_0} \diff s_0\, \int\diff^3 s \,  \e^{\Ci ps} \pdiff{X_0} F(X,s) \nonumber \\
    &=& \pdiff{X_0} \tilde  F(X,p) -2\Big[\e^{2\Ci p_0 X_0}F(X, s_0=2X_0, \F p) \nonumber \\
    && ~~~~~~~+ \e^{-2\Ci p_0 X_0}F(X, s_0=-2X_0, \F p)\Big].
\label{eq:TE:limits1}
  \end{eqnarray}
Assuming $\lim_{s_0\toward\infty}F(s_0) = 0 = \lim_{s_0\toward\infty}\rho(s_0)$ yields
$
    \lim_{X_0\toward\infty}\pDiff{X_0}=\pdiff{X_0}.
$
The leading-order transport equations result as
  \begin{eqnarray}
\label{eq:TE:motionFWigner_LO}
      \pDiff{X_0}[\tilde F(X,p)]
      &=& \Ci\sigma_2 
         \Big\lbrace
      \tilde F(X, p)\tilde\Sigma^\rho(X, p) - \tilde \rho(X, p) \tilde \Sigma^F(X,p) \Big\rbrace 
	   \\
\label{eq:TE:motionRhoWigner_LO}
      \pDiff{X_0}[\tilde \rho(X,p)]
     &=&\,0.
  \end{eqnarray} 
  The contributions from the integration limits account for the fact that the correlation functions are initialized at some initial time $x_0=y_0=0$. These contributions can be removed for sufficiently late times as the time correlations are expected to vanish for sufficiently large relative times.

To next-to-leading order in the gradient expansion, the transport equations read
  \begin{eqnarray}
\label{eq:TE:motionFWigner_NLO}
    \pDiff{X_0}[\tilde F] 
    &=&\,\mathrm{LO} +\sigma_2\Bigg\lbrace \big[\pdiff{X_0} \Sigma^{(0)}(X)\big] \pdiff{p_0} \tilde F(X,p) \nonumber\\
    && - \frac 1 2\Big[
        \lbrace 
          \tilde\Sigma^+(X,p),\tilde F(X,p) 
        \rbrace_0
       +\lbrace 
          \tilde \Sigma^F(X,p),\tilde \rho^+(X,p)
        \rbrace_0 \Big]\Bigg\rbrace,
    \\
    \pDiff{X_0}[\tilde\rho] 
    &=&\, \sigma_2\Bigg\lbrace \big[\pdiff{X_0} \Sigma^{(0)}(X)\big] \pdiff{p_0} \tilde\rho(X,p) \nonumber\\
    && - \frac 1 2\Big[
        \lbrace 
          \tilde\Sigma^+(X,p),\tilde\rho(X,p) 
        \rbrace_0
       +\lbrace 
          \tilde \Sigma^\rho(X,p),\tilde \rho^+(X,p)
        \rbrace_0 \Big]\Bigg\rbrace,
  \end{eqnarray}
where we introduced the Poisson brackets with respect to $X_0$ and $p_0$:
  \begin{equation}
    \lbrace\tilde A,\tilde B\rbrace_0 = \pdiffer{\tilde A}{p_0}\pdiffer{\tilde B}{X_0}-\pdiffer{\tilde A}{X_0}\pdiffer{\tilde B}{p_0}.
  \end{equation} 
%

\section{Comparison of dynamical and transport equations}
\label{sec:DynvsTrans}
In the following we compare the transport equations derived in the preceding section, in leading order, Eq.~(\eqref{eq:TE:motionFWigner_LO}), and next-to-leading order, Eq.~(\eqref{eq:TE:motionFWigner_NLO}), of the gradient expansion, with the equations of motion (\eqref{eq:TE:motionF1}).
This is achieved in two steps. 
First, we calculate $F$ and $\rho$ using equations (\eqref{eq:TE:motionF1}), (\eqref{eq:TE:motionRho1}) in the NLO $1/\mathcal N$-approximation. 
We then take the results of this calculation and, after a Fourier transformation to Wigner space, we calculate both the left-hand and the right-hand sides of equations (\eqref{eq:TE:motionFWigner_LO}) and (\eqref{eq:TE:motionFWigner_NLO}). 
The left-hand side yields the time derivatives of the solutions of the 2PI dynamic equations with respect to the centre time coordinate, while the right-hand side generates the corresponding approximative derivatives in leading order and next-to-leading order of the gradient expansion.

\subsection{Physical setup}
We consider a dilute homogeneous one-dimensional ($D=1$) gas of spinless sodium atoms with mass $m=22.99\,\unit u$ ($^{23}\text{Na}$) confined in a periodic box of length $L=N_s a_s\approx 43\,\unit{\mu m}$.
The spacing of the numerical grid is $a_s=1.33\,\unit{\mu m}$, and the number of lattice points is $N_s=32$. 
This results in discrete momentum modes\footnote{Since the symbols $x$, $p$ etc. so far have been defined as $x=(x_0,\F x)$, $p=(p_0,\F p)$, we retain the boldface notation for space-like quantities although we consider the one-dimensional case.} $\F p=(2N_s/L)\sin(n_{p}\pi/L)$, with $n_{p}=-N_s/2...N_s/2$. 
Using different initial line densities $n_1$ between $10^5\,\unit m^{-1}$ and $10^7\,\unit m^{-1}$ the total particle number varies in the range from $\sim4$ to $\sim400$ particles. 

For our investigations we consider different interaction strengths characterized by the parameter $\gamma$ which is related to the 1D coupling constant as
  \begin{equation}
    g = \frac{\gamma n_1}m.
\label{eq:NR:defGamma}
  \end{equation} 
  We vary $\gamma$ in three steps from weak to strong interactions, $\gamma=1.5\cdot10^{-3}$, $0.15$, $15$, thereby keeping the quantity $\gamma n_1^2=mgn_1$ fixed by choosing the line densities given above. 
In this way, the effect of quantum fluctuations is increased with $\gamma$ while the dynamics in the classical statistical approximation remains unchanged \cite{Berges2007a}.

The spatial homogeneity allows for a Fourier transformation of the dynamic equations (\eqref{eq:TE:motionF1}, \eqref{eq:TE:motionRho1}) with respect to the spatial relative coordinate $\F s$ to relative-momentum space:
  \begin{eqnarray}
    &&\pdiff{x_0}F(x_0,y_0,\F p) 
    =\, \Ci\sigma_2\Bigg\lbrace M(x_0,\F p) F(x_0,y_0,\F p) 
    \nonumber\\
    && ~~~ +\int\limits_{0}^{x_0}\diff z_0\, \Sigma^\rho(x_0,z_0,\F p)F(z_0,y_0,\F p)-\int\limits_{0}^{y_0} \diff z_0\, \Sigma^F(x_0,z_0,\F p)\rho(z_0,y_0,\F p)\Bigg\rbrace,
  \label{eq:DE:Fp}
    \\ 
    &&\pdiff{x_0}\rho(x_0,y_0,\F p) 
    =\, \Ci\sigma_2\Bigg\lbrace M(x_0, \F p) \rho(x_0,y_0,\F p) 
    \nonumber\\
    && ~~~ +\int\limits_{y_0}^{x_0}\diff z_0\, \Sigma^\rho(x_0,z_0,\F p)\rho(z_0,y_0,\F p)\Bigg\rbrace,
  \label{eq:DE:rhop}
  \end{eqnarray} 
  where the generalized energy matrix is given by
  \begin{equation}
    M(x_0, \F p) 
    = \frac{\F p^2}{2m}\opUnit 
    + \frac g 2\int_{\F k}\big[\trace\lbrace F(x_0,x_0,\F k)\rbrace\opUnit
    + 2F(x_0,x_0,\F k)\big],
  \end{equation} 
according to (\eqref{eq:DE:Sigma0Def}) and \cite{Gasenzer:2005ze}. 
Here, $\int_{\F k}\equiv(2\pi)^{-1}\int\diff\F k$ denotes the one-dimensional momentum-integral, and the self-energy contributions are obtained, in NLO of the $1/\cal N$ expansion, from the Fourier transforms of Eqs.~(\ref{eq:DE:SigmaF}) to (\ref{eq:DE:IRho}) which are provided in \ref{app:SelfEn1Dhom}, cf.~Eqs.~(\ref{eq:SigmabarFpofFrho})--(\ref{eq:Irhop}).  

  With initial values for $F(0,0,\F p)$ and $\rho(0,0,\F p)$, the above coupled system of integro-differential equations yields the time evolution of the correlation functions. 
To fix $F(0,0,\F p)$, Eq.~(\ref{eq:IniCondF}), we chose a Gaussian momentum distribution peaked around $\mathbf{p}=0$, with a width of $\sigma=6.5\cdot 10^4\, \unit m^{-1}$: $f_{0}(\F p) = n(0,\F p)+\frac 1 2$, where
  \begin{equation}
    n(0,\F p)=\frac {n_1}{\sqrt\pi \sigma}\e^{-\F p^2/\sigma^2}.
    \label{eq:npini}
  \end{equation} 
%
\begin{figure}[tb]
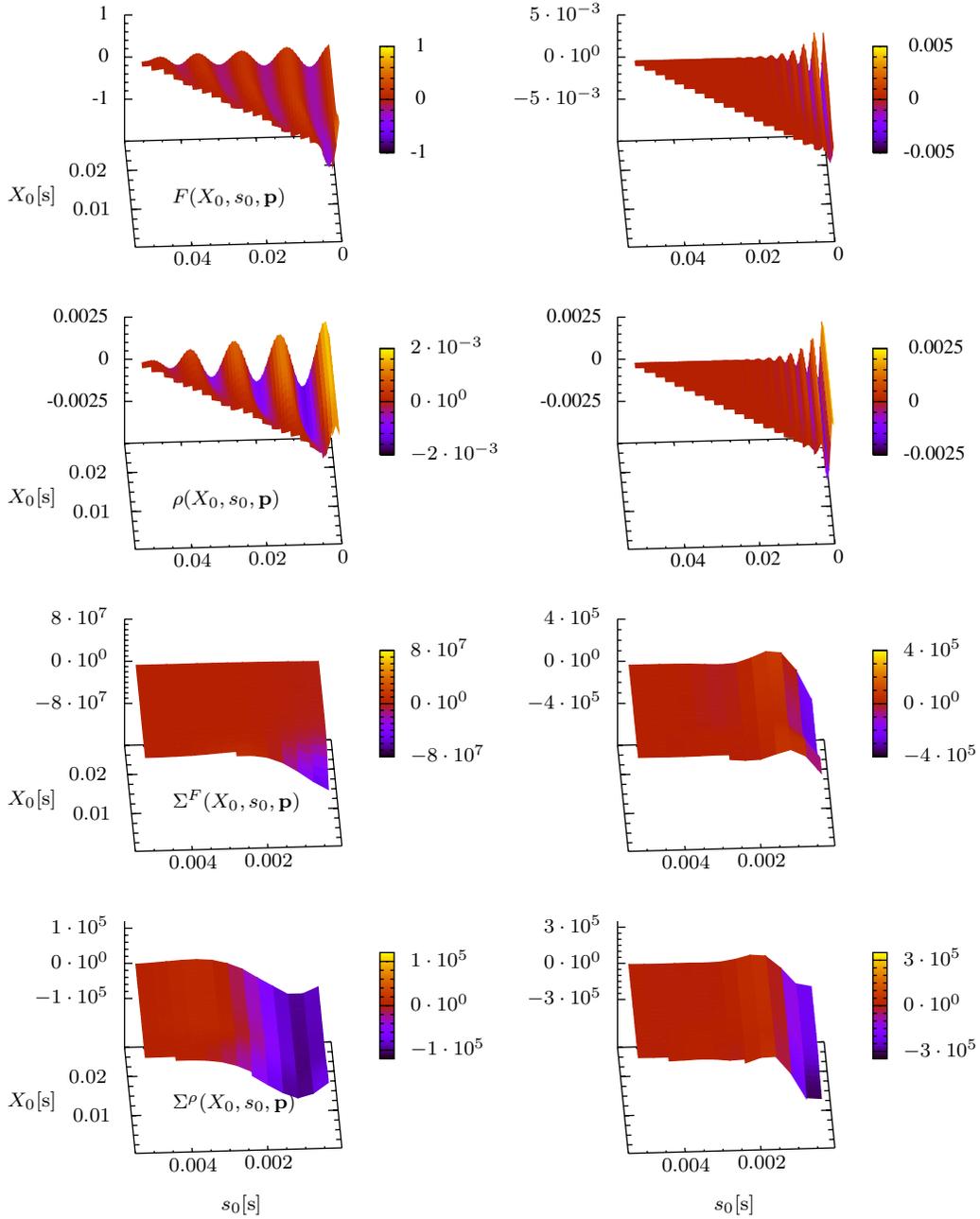

\begin{flushright}
     \ \\[1ex]
      \begin{scriptsize}
     \include{./fig2}
      \end{scriptsize}
    \caption[]{The statistical and spectral two-point functions and self-energies for two different momentum modes as a function of the relative and centre-of-time coordinates $s_0$ and $X_0$, respectively.
Left column: momentum mode $n_p=0$, right column: $n_p=8$. 
From top to bottom: $F$, $\rho$, $\Sigma^F$, and $\Sigma^\rho$. 
The figures illustrate that all correlation functions vanish for sufficiently large relative times. 
See the main text for the parameters chosen.
}
\label{fig:NR:Xs3d}
\end{flushright}
\end{figure}

The Bose commutation relations fix, as in Eq.~(\ref{eq:IniCondRho}), the initial values for the spectral function: $\rho(x_0,x_0,\F p) = -\Ci\sigma_2$.

\subsection{Numerical solution of the dynamic equations}
The dynamic equations (\ref{eq:DE:Fp}), (\ref{eq:DE:rhop}) were solved using the techniques described in Refs.~\cite{Gasenzer:2005ze,Berges2007a}.
From their solutions, the two-point functions in Wigner space were obtained by a discrete Fourier transform along the relative time direction $s_0$.

We expect the transport equations to reproduce, to a good approximation, the full dynamical evolution, if the two-point functions vary slowly with respect to the centre of time coordinate $X_0$ as compared to the change with the relative time coordinate $s_0$. 
Figure \ref{fig:NR:Xs3d} shows $F(X_0, s_0, \F p)$ for two different momentum modes, together with the corresponding evolution of $\rho$, $\Sigma^F$  and $\Sigma^\rho$. 
We see that for sufficiently late times the functions fall off to zero with increasing relative time $s_0$. 
The rapid decay of the self-energy as compared to $F$ and $\rho$ can be understood by noting that, in the next-to-leading order $1/\mathcal N$ expressions, see Eqs.~(\eqref{eq:DE:SigmaF}) and 
 (\eqref{eq:DE:SigmaRho}) as well as (\eqref{eq:DE:IF}) and (\eqref{eq:DE:IRho}), the correlators $F$ and $\rho$ enter to the third power. 
Compared to their oscillations along the relative time direction, we note only a weak dependence on the centre-of-time coordinate.
The form of the correlators in the temporal plane shown in Fig.~\ref{fig:NR:Xs3d} is to be compared with that of the correlators for an ideal gas which show an undamped oscillatory dependence in the $s_0$ direction, with the frequency given by the free dispersion $p_0={\F p}^2/2m$, see Eqs.~(\ref{eq:TE:ref1}), (\ref{eq:TE:ref2}). 

\begin{figure}[tb]
    \begin{center}
     \ \\[1ex]
    \begin{scriptsize}
      \includegraphics[width=0.7\textwidth]{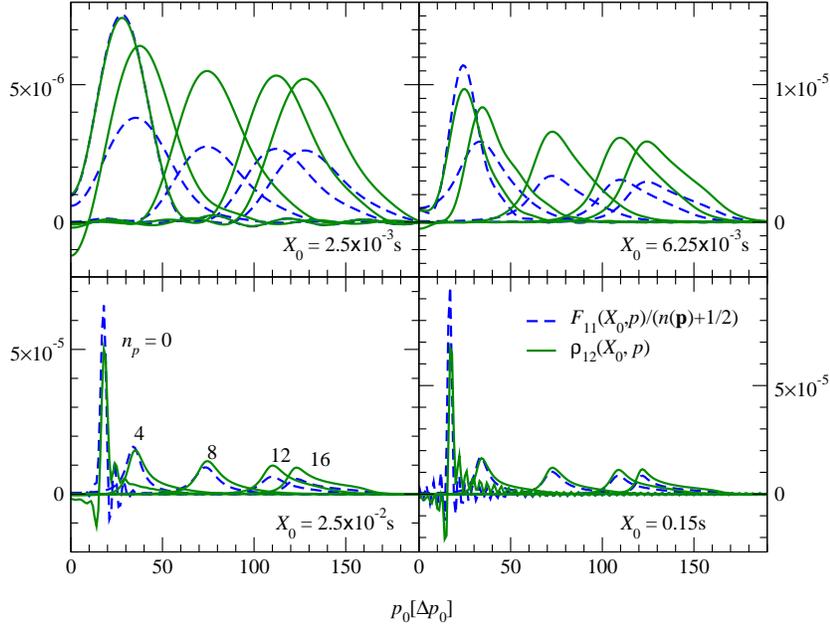}
      \ \\
    \end{scriptsize}
    \caption[]{(color online) Spectral function $\tilde\rho$ (green solid lines), and statistical function $\tilde F$, normalized to the occupation number of the momentum mode $\F p$ (blue dashed lines), as functions of $p_0$, for different times $X_0$ and different momentum modes $n_p$. 
The difference between the respective spectral and normalized statistical functions decreases, indicating the emergence of a fluctuation dissipation relation.
The good correspondence for the zero mode $n_p=0$ at the initial time is related to the choice of the initial condition.
}
\label{fig:NR:fluctdisprel}
\end{center}
\end{figure}

As in the case of an ideal gas discussed in \ref{app:NonintGas}, the two-point functions $\tilde F$ and $\tilde\rho$ of an interacting system are expected to be related by a fluctuation dissipation relation,
  \begin{equation}
    \tilde F(X,p) 
    = \Ci\sigma_2\Big(n(\F p)+\frac 1 2\Big) \tilde\rho(X,p),
\label{eq:TE:fluctuationdissipation2}
  \end{equation} 
if the system is close to equilibrium \cite{KadanoffBaym1962a,Aarts:1997kp}. 
In  Figure \ref{fig:NR:fluctdisprel} we explicitly show that the function $\tilde\rho_{21}(X_0,p)$ approaches $\tilde F_{11}(X_0,p)/(n(X_0, \F p)+\frac 1 2)$ during the time evolution described by the 2PI dynamic equations (\ref{eq:DE:Fp}), (\ref{eq:DE:rhop}), where $n(X_0, \F p)= F_{11}(X_0, s_0=0, \F p) -\frac 1 2$.
%
\begin{figure}[tb]
\begin{center}
\includegraphics[width=0.6\textwidth]{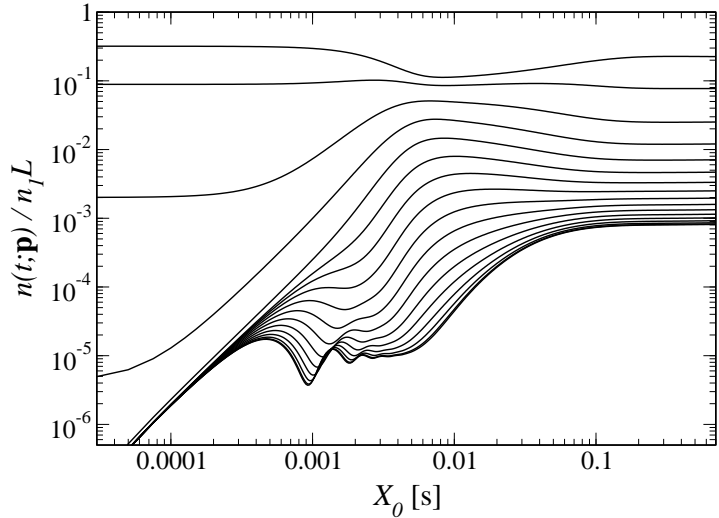}
\end{center}
\vspace*{-3ex}
\caption{
The momentum-mode occupation numbers $n(t,{\F p})/n_1L$ normalized by the total number of atoms in the box, $n_1L=853$, as functions of time. 
The gas is in a far-from-equilibrium state initially, characterized by a Gaussian distribution $n(0,{\F p})$, Eq.~(\protect\ref{eq:npini}), with width
$\sigma=1.3\cdot10^5$m$^{-1}$. 
It is weakly interacting, $\gamma=1.5\cdot10^{-3}$.
Shown are the populations of the modes with ${\F p}=2N_s/L\sin(n_p\pi/N_s)$, $n_p=0$ (uppermost curve), $2,4,...,N_s/2$ (sequentially underneath), for $N_s=64$, and one has $n(t;-{\F p})=n(t;{\F p})$. 
A fast short-time dephasing period is followed by a long quasistationary drift to the final equilibrium distribution. 
Notice the double-logarithmic scale.
}
\label{fig:npoft}
\end{figure}

\subsection{2PI dynamical versus transport equations}
\label{sec:NR:em_vs_te}
\begin{figure}[tb]
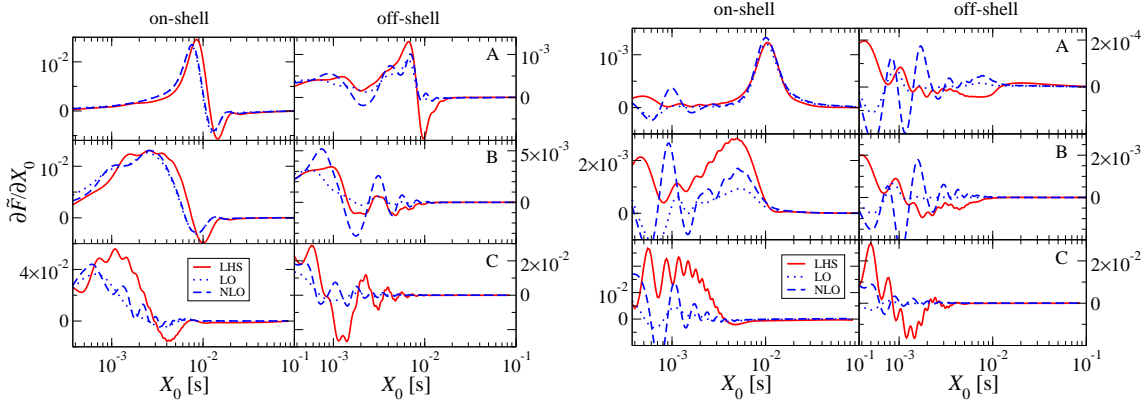

    \begin{center}
     \ \\[1ex]
    \begin{scriptsize}
      \includegraphics[width=0.48\textwidth]{./fig5a.eps}
      \includegraphics[width=0.48\textwidth]{./fig5b.eps}
      \ \\
    \end{scriptsize}
    \caption[]{(Color online) 
Time-derivative of the statistical two-point function as a function of centre time $X_0$ for two different momentum modes, $n_p=4$ (left panel) and $n_p=8$ (right panel): 
Comparison of results from 2PI dynamic equations and those from their kinetic approximation.
The (red) solid lines correspond to the left-hand side (LHS), the (blue) dashed(-dotted) lines to the the right-hand side in leading order (LO) (next-to-leading order, NLO) of the gradient expansion, Eqns. \eq{eq:TE:motionFWigner_LO} and \eq{eq:TE:motionFWigner_NLO}, respectively.
In the left column of each panel, the frequency $p_0$ is that of the peak of the spectral function, cf. Fig.~\ref{fig:NR:fluctdisprel} (`on-shell'), in the respective right columns, $p_0$ has been chosen two half-widths away from it (`off-shell'). 
From top to bottom, the line density $n_1$ is rescaled as well as the interaction parameter $\gamma$, so that $m g n_1=\gamma n_1^2$ is kept fixed: 
Top row: case A $n_1=10^7 \unit m^{-1}$, $\gamma=1.5\cdot 10^{-3}$; 
middle row: case B $n_1=10^6 \unit m^{-1}$, $\gamma=0.15$; 
bottom row: case C $n_1=10^5 \unit m^{-1}$, $\gamma=15$. 
}
\label{fig:NR}
\end{center}
\end{figure}

We compare the time evolution as derived from the 2PI dynamic equations to their kinetic approximations following the procedure stated at the beginning of the section.  
The absolute values of the left-hand side (LHS), the right-hand side in leading order (LO) and the right-hand side in next-to-leading order (NLO) of Eqns. (\eqref{eq:TE:motionFWigner_LO}, \eqref{eq:TE:motionFWigner_NLO}) are drawn in Figure \ref{fig:NR}.

Three generic time regimes are found \cite{Berges:2001fi}: 
Strong oscillations characteristic for early times, slow drifting at intermediate times, and a late-time approach to equilibrium characterized by vanishing time derivatives.
These time regimes can already be seen in the corresponding time evolution of momentum-mode occupation numbers $n(X_0, \F p)= F_{11}(X_0, s_0=0, \F p) -\frac 1 2$ shown in Fig.~\ref{fig:npoft} for a corresponding system with $N_s=64$. 
Cf.~Ref. \cite{Berges2007a} for a more detailed discussion of the numerical evaluation of the 2PI dynamic equations.
 
The different regimes can be understood as follows:
Close to the initial time, the strong oscillations of $F(X_0, s_0,p)$ and $\rho(X_0, s_0, p)$ in the centre time $X_0$ are due to the finite integration limits $s_0=\pm2X_0$ in the $s_0$ direction.
Consider the analytic solutions for the interaction free case, Eqs.~(\eqref{eq:TE:ref1}), (\eqref{eq:TE:ref2}). 
Here, the oscillation frequency scales $\propto p^2$ while the amplitude is constant, such that correlations do not decay. 
Consequently, the corresponding quantities in Wigner space, $\tilde F(X_0, p_0, \F p)$ and $\tilde \rho(X_0, p_0, \F p)$, show the same oscillations if we restrict the Fourier transformation to the finite range $-2X_0<s_0<2X_0$.
Besides these oscillations the amplitude of the correlation functions, in the interacting case, also decays with increasing $s_0$.

The intermediate drifting regime is reached (for $\gamma=1.5\cdot10^{-3}$ at $X_0\simeq0.003\,$s) when the contributions of the integration limits can be neglected.
Non-vanishing derivatives of $\tilde F$ and $\tilde\rho$ with respect to $X_0$ now solely result from the evolution of the correlations $F(X_0, s_0, \F p)$ and $\rho(X_0, s_0, \F p)$ in $X_0$.
As can be seen in Figure \ref{fig:NR:Xs3d}, the change with $X_0$ is slow compared to that with $s_0$. 
Finally, equilibrium is approached when all time derivatives vanish.

We consider in more detail the cases corresponding to (A) weak, (B) moderate, and (C) strong effective interactions between the atoms:

\paragraph{Case A}
corresponds to weak interactions, $\gamma=1.5\cdot 10^{-3}$, i.e., to an essentially classical statistical evolution \cite{Berges2007a}. 
The evolution is expected to be well described by Boltzmann-type equations after the initial oscillations have damped out. 
The initial line density is set to $n_1=10^7$, which corresponds to $\sim420$ particles. 
Figure \ref{fig:NR}A (top row) shows the left-hand side (LHS, solid line) as well as the right-hand side of the transport equations in leading order (LO, dashed line) and next-to-leading order (NLO, dashed-dotted line) of the gradient expansion for two different momentum modes $n_p=4$ (left panel) and $n_p=8$ (right panel).
In the respective left columns, the frequency $p_0$ is that of the peak of the spectral function, cf. Fig.~\ref{fig:NR:fluctdisprel} (`on-shell'), in the right columns, $p_0$ has been chosen two half-widths away from it (`off-shell'). 
We find that the 2PI dynamic and the kinetic equations in general give the same results only as soon as the occupation numbers do not change any longer at all.
\\

The next-to-leading order contributions depend on the $p_0$ derivatives of the correlation functions. Compared to the interaction free case (\eqref{eq:TE:analyticalRho}), the correlation functions in Wigner representation become smooth functions of $p_0$ as can be seen in Figure \ref{fig:NR:fluctdisprel}. 
This results in continuous $p_0$-derivatives, and therefore in non-vanishing contributions in next-to-leading order. 
However, as shown by the red dashed line, no significant contributions to the transport equations are found after the decay of the initial oscillations. 
\begin{figure}[tb]
    \begin{center}
     \ \\[1ex]
    \begin{scriptsize}
      \includegraphics[width=0.7\textwidth]{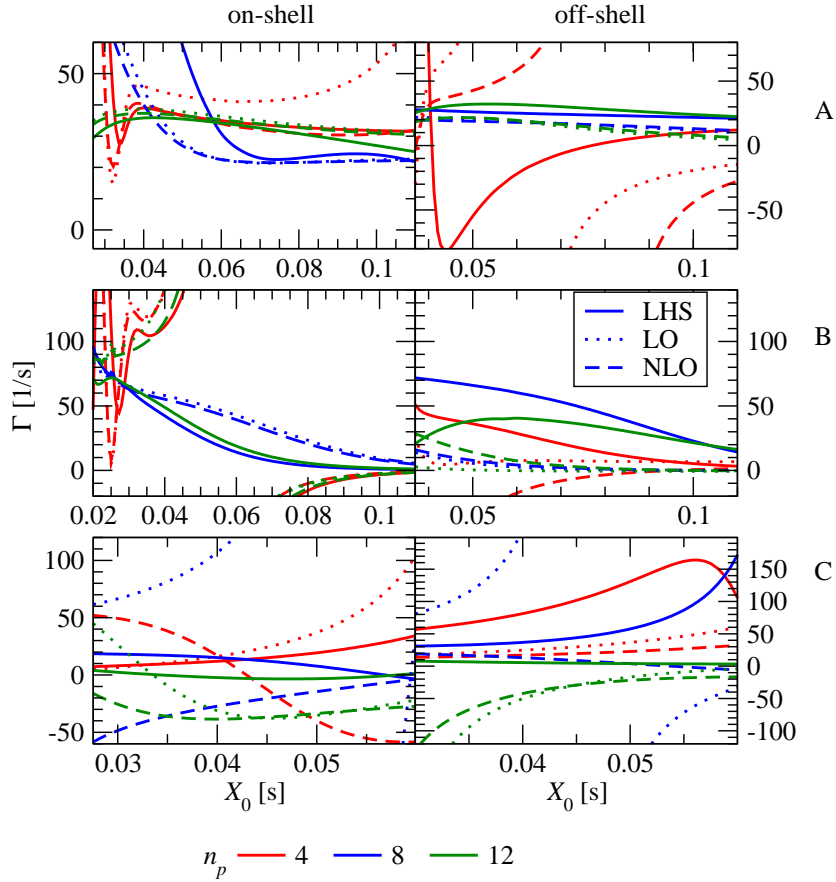}
      \ \\
    \end{scriptsize}
    \caption[]{(Color online) The figure shows the time dependent decay constant $\Gamma$ as defined in Eq. (\eqref{eq:TE:decayconst}), where the time derivative $\diff \tilde F/\diff X_0$ is replaced by the left-hand side (LHS, solid line), the right-hand side in leading order (LO, dotted line) and in next-to-leading order (NLO, dashed line) of the gradient expansion, Eqns. (\eqref{eq:TE:motionFWigner_LO}, \eqref{eq:TE:motionFWigner_NLO}), for the on-shell and the off-shell case as a function of the centre time. 
From top to bottom, the line density $n_1$ is rescaled as well as the interaction parameter $\gamma$ with $\gamma n_1^2=\const$, compare Fig. \ref{fig:NR}.
The three rows correspond to cases A--C as described in the text.
Colors indicate the momentum mode: $n_p=4$ (red), $n_p=8$ (blue), $n_p=12$ (green). 
Non-constant (and negative) values for $\Gamma$ result from the fact that the time regime of exponential approach to equilibrium is not yet reached. 
}
\label{fig:decayconst}
\end{center}
\end{figure}

\paragraph{Case B} 
We increase the interaction parameter to $\gamma=0.15$ while the initial line density is decreased to $n_1=10^6$, corresponding to a total number of $\sim 40$ particles.
Note that in this way $g\propto\gamma n_1$ increases by a factor of $10$ such that quantum statistical correlations grow in importance, see Ref.~\cite{Berges2007a}. 
Our results are shown in the second row of Fig.~\ref{fig:NR}. 
We find equilibrium to be reached faster, in particular for the higher momentum modes, while the intermediate drifting regime observed in case A is reduced. 
For the lower momentum modes we get similar results as in case A while for the higher momentum modes there is an essential difference to the preceding case: correspondence between 2PI and transport equations is not reached until equilibration occurs. 

\paragraph{Case C.} 
We finally choose strong coupling, $\gamma=15$, and decrease the line density to $n_1=10^5\unit m^{-1}$ corresponding to a total number of $4$ particles. 
Our results are shown in the third line of Fig.~\ref{fig:NR}.
With this, we find qualitatively similar results as in the preceding cases and a faster approach to an equilibrium configuration.

To study in more detail the late-time behaviour, we assume that, at late times, the statistical correlation function decays exponentially to its equilibrium value with a decay constant $\Gamma(p)$,
\begin{equation}
  \tilde F(X_0, p)\approx \tilde F(X_0=\infty, p) + \Delta\tilde F(p)\e^{-\Gamma(p) X_0},
\end{equation}
where $\Delta\tilde F(p)$ is some constant independent of $X_0$.
In order to be independent of the equilibrium value, we plot, in Fig.~\ref{fig:decayconst}, the time-dependent expression
\begin{equation}
  \Gamma(X_0, p) 
  = - \frac{\partial^2 \tilde F(X_0, p)/\partial X_0^2}
      {\partial \tilde F(X_0, p)/\partial X_0},
\label{eq:TE:decayconst}
\end{equation} 
where $\partial \tilde F/\partial X_0$ is given by the LHS, 
as well as by the 
LO and 
NLO  expressions on the RHS of 
Eq.~\eq{eq:TE:motionFWigner_LO}, for three different momentum modes.
The graphs correspond to the same choice of parameters and momenta as in Fig.~\ref{fig:NR}. 
We focus the range of times $X_0$ to those where $\Gamma(X_0, p)$ is settling to a constant, indicating the emergence of an evolution according to kinetic theory.
The top row of graphs in Fig.~\ref{fig:decayconst} shows that the different mode evolutions are settling to an exponential decay at times between $0.07$ and $0.09$ seconds, i.e., when compared with Fig.~\ref{fig:npoft}, only when there are almost no changes seen any more in the occupation number evolution.
Hence, despite the fact, that a fluctuation dissipation relation is established almost an order of magnitude in time earlier, the kinetic approximation becomes strictly valid only at very late times.
Fig.~\ref{fig:decayconst} also shows that it is in general not sufficient to take into account the LO approximation in the gradient expansion only.

Our findings become even more pronounced in cases B and C, when the system becomes more strongly correlated.
For the largest couplings (case C) even to NLO in the gradient expansion the non-Markovian results of the 2PI dynamics are recovered.

\begin{figure}[tb]
    \begin{center}
     \ \\[1ex]
    \begin{scriptsize}
      \includegraphics[width=0.7\textwidth]{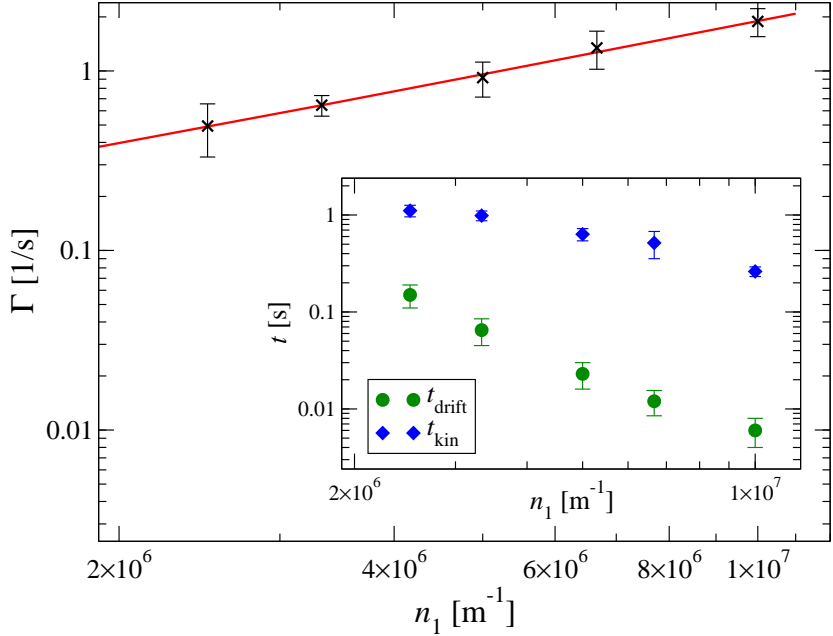}
      \ \\
    \end{scriptsize}
    \caption[]{(Color online) The decay constant $\Gamma$ as defined in Eq. (\eqref{eq:TE:decayconst}) as a function of the particle density $n_1$, extracted from the solution of the 2PI dynamic equations at times $t_\mathrm{kin}$ as indicated with (blue) diamonds in the inset figure, 
after which the evolution can be described to a good approximation by an exponential decay.
In the inset, the corresponding times $t_\mathrm{drift}$ at which the slow drift of the occupation number sets in are indicated with (green) circles.
The error bars indicate the variation of $\Gamma$, $t_\mathrm{kin}$, and $t_\mathrm{drift}$ over the different momentum modes.
The red line is a linear fit.
}
\label{fig:Gammaofn1}
\end{center}
\end{figure}

We finally study the dependence of the decay constants $\Gamma$ on the density $n_1$ of particles along the one-dimensional system.
Fig.~\ref{fig:Gammaofn1} shows $\Gamma$, extracted at times $t_\mathrm{kin}$ indicated in the inset, for five different densities $n_1$.
Error bars indicate the variation over the different momentum modes.
The times $t_\mathrm{kin}$ have been chosen such that $\Gamma(t)$ is found to remain approximately constant for times $t\ge t_\mathrm{kin}$.
We find an approximately linear dependence of $\Gamma$ on $n_1$ which indicates that the source of damping is rather an off-shell two-body than a three-body scattering effect.

The inset of Fig.~\ref{fig:Gammaofn1} also shows the times at which the fluctuation dissipation relation between the statistical and spectral correlation functions starts to hold.
The difference between this time and the respective $t_\mathrm{kin}$ increases with growing $n_1$ and therefore with increasing effective interaction strength.

\section{Conclusions and outlook}
\label{sec:Concl}
We have studied the far-from-equilibrium dynamics of an ultracold, one-dimensional Bose gas and focused on the comparison between a fully dynamical approach on the basis of the 2PI effective action and the corresponding kinetic approximation in the form of Boltzmann-type transport equations. 

The 2PI effective action which allowed to derive dynamic equations by use of Hamilton's principle was considered in next-to-leading order of an expansion in inverse powers $1/\cal N$ of the number of field components. 
This approximation does not rely on a small coupling constant as expansion parameter and therefore is non-perturbative.
Dynamic equations derived from the 2PI effective action automatically preserve essential conserved quantities as the total particle number and the total energy of the system. 
  
The transport equations were derived from the 2PI dynamic equations for the two-point correlation functions by means of a gradient expansion with respect to centre coordinates and a subsequent Fourier transformation with respect to relative coordinates, i.e., a Wigner transformation. 
Furthermore, the details of the initial state are neglected by sending the initial time to minus infinity. 
This approximation relies on the fact that the time evolution of the correlation functions with respect to centre time occurs relatively slowly as compared to the oscillations with respect to relative time. 

\bigskip
We considered a homogeneous one-dimensional system of spinless bosonic atoms with an initially Gaussian momentum mode distribution.
Both, weakly and strongly interacting systems were considered, such that the significance of quantum fluctuations to the time evolution could be studied.
A comparison of the statistical and spectral correlation functions allowed to determine the time scale at which they are connected by a fluctuation dissipation relation.
This timescale was shown to coincide with the time scale at which the transport equations set in to represent a valid description of the dynamical evolution.

For weak couplings we observe good correspondence of dynamic and transport equations after an initial period of oscillations. 
However, off-shell effects are not covered by transport equations before equilibration occurs. 
Increasing the dimensionless interaction strength parameter $\gamma$ led to significant differences between the 2PI dynamic and the transport equations. 
Our results show that the kinetic description of the time evolution in terms of an exponential decay with decay rate $\Gamma$ generically sets in to be valid only at very large times, when no essential change in the momentum profile of the system occurs any more.
Moreover, they indicate that the late-time evolution is predominantly due to two-body off-shell scattering effects.

\section*{Acknowledgments}
\noindent
We are very grateful to J\"urgen Berges, Ana Maria Rey, and Kristan Temme for valuable discussions, and to Werner Wetzel for his continuing support concerning computing facilities.
This work has been supported by the Deutsche Forschungsgemeinschaft (T.G.).

\begin{appendix}

\section{Transport equations in Wigner relative time directions}
\label{app:DEwrts0}
\subsection{Transport equations}
\label{app:TEins0}
\noindent
In Section \ref{sec:TransportEq} we have derived transport equations for the evolution along the Wigner centre time $X_0$ by adding the dynamic equations governing $\partial_{x_0}F(x,y)$ and $\partial_{y_0}F(x,y)$, respectively, cf. Eqs.~(\ref{eq:TE:FAbs}) and (\ref{eq:TE:rhoAbs}).
A second set of equations for the derivatives of $F(X,s)$ and $\rho(X,s)$ with respect to $s_0$ results when subtracting the respective expressions:
  \begin{eqnarray}
    \pdiff{s_0}F(X,s) 
    &=& \frac{\Ci\sigma_2}2
      \Big\lbrace [\opHoneB(X+\frac s 2)+\opHoneB(X-\frac s 2)
    \nonumber \\
    && ~~~~~~~~~~ +\Sigma^{(0)}(X+\frac s 2)+\Sigma^{(0)}(X-\frac s 2)] F(X,s) \nonumber \\
    && +\int_{s'} \theta(X_0+s_0'-\frac{s_0}2) \big[ \Sigma^R(X+\frac {s'} 2 , s-s') F(X-\frac{s-s'}2, s') 
     \nonumber \\
     && ~~~~~~~~~~ + G^R(X+\frac{s'}2, s-s') \Sigma^F(X-\frac{s-s'}2,s') \nonumber \\
     && ~~~~~~~~~~ +\Sigma^F(X+\frac {s'} 2 , s-s') G^A(X-\frac{s-s'}2, s') \nonumber \\
     && ~~~~~~~~~~ + F(X+\frac{s'}2, s-s') \Sigma^A(X-\frac{s-s'}2,s')\big]\Big\rbrace,
\label{eq:TE:FRel}
  \end{eqnarray}
  \begin{eqnarray}
    \pdiff{s_0}\rho(X,s) 
     &=& \frac{\Ci\sigma_2}2 \Big\lbrace 
       \big[\opHoneB(X+\frac s 2)+\opHoneB(X-\frac s 2) 
     \nonumber \\
     && ~~~~~~~~~~ +\Sigma^{(0)}(X+\frac s 2) +\Sigma^{(0)}(X-\frac s 2) \big]\rho(X,s) \nonumber \\
     && +\int_{s'}\big[\Sigma^R(X+\frac {s'}2,s-s') \rho(X-\frac{s-s'}2,s') \nonumber \\
     && ~~~~~~~~~~ +\rho(X+\frac{s'}2,s-s')\Sigma^A(X-\frac{s-s'}2,s') \nonumber \\
     && ~~~~~~~~~~ +\Sigma^\rho(X+\frac{s'}2,s-s')G^A(X-\frac{s-s'}2, s') \nonumber \\ 
     && ~~~~~~~~~~ +G^R(X+\frac{s'}2,s-s')\Sigma^\rho(X-\frac{s-s'}2, s')
     \big]
    \Big\rbrace.
\label{eq:TE:rhoRel}
  \end{eqnarray} 
In analogy to the derivation of the gradient expansion with respect to centre coordinates $X$ in Section \ref{sec:GradExp} one obtains LO transport equations in the relative time $s_0$,
  \begin{eqnarray}
   &&\pdiff{s_0}F(X,s) 
   = {\Ci\sigma_2}
         \Bigg\lbrace
           \left[-\frac {\laplace_{\F s}} {2m}+\Sigma^{(0)}(X)\right]F(X,s)
	\nonumber \\
    && ~~~~~~~~ +\frac 1 2\int_{s'}
       \big[\Sigma^+(X,s-s')F(X,s')+\rho^+(X,s-s')\Sigma^F(X,s')\big]\Bigg\rbrace,
\label{eq:TE:motionF_LO2}
   \\
   &&\pdiff{s_0}\rho(X,s) 
   = {\Ci\sigma_2}
         \Bigg\lbrace
           \left[-\frac {\laplace_{\F s}} {2m}+\Sigma^{(0)}(X)\right]\rho(X,s)
	   \nonumber \\
    && ~~~~~~~~ +\frac 1 2\int_{s'}
       \big[\Sigma^+(X,s-s')\rho(X,s')+\rho^+(X,s-s')\Sigma^\rho(X,s')\big]\Bigg\rbrace,
\label{eq:TE:motionRho_LO2}
  \end{eqnarray} 
and their NLO corrections as
  \begin{eqnarray}
   \lefteqn{\pdiff{s_0}F(X,s)=\mathrm{LO}} \nonumber\\  
   &&+\frac {\Ci\sigma_2} 4 \int_{s'}\Big [[s'_0\pdiff{X_0}\Sigma^\rho(X, s-s')]F(X,s')
     -\Sigma^\rho(X, s')[s'_0\pdiff{X_0} F(X,s-s')] \nonumber \\
   && ~~ +[s'_0\pdiff{X_0} \rho(X,s-s')]\Sigma^F(X,s') 
           - \rho(X,s')[s'_0\pdiff{X_0} \Sigma^F(X,s-s')]\Big], 
\label{eq:TE:motionF_NLO2}
   \\
   \lefteqn{\pdiff{s_0}\rho(X,s)=\mathrm{LO}.} 
\label{eq:TE:motionRho_NLO2}
  \end{eqnarray} 
Here, LO denotes the respective right hand side of Eqs.~(\ref{eq:TE:motionF_LO2}) and (\ref{eq:TE:motionRho_LO2}).
The functions $\rho^+$ and $\Sigma^+$ are defined in Eqs.~(\ref{eq:rhoplus}) and (\ref{eq:Sigmaplus}), respectively.

Proceeding with the transformation to Wigner space as in Sect.~\ref{sec:TransWigner} we obtain the LO,
  \begin{eqnarray}
    \pDiff{p_0}[\tilde F(X,p)]
    &=&\, \Ci\sigma_2\Bigg\lbrace\Big[ \frac {\F p^2}{2m}+\Sigma^{(0)}(X)\Big] \tilde F(X,p) 
    \nonumber\\
    && + \frac 1 2 \Big[\tilde \Sigma^+(X,p)\tilde F(X,p)+\tilde \rho^+(X,p)\tilde\Sigma^F(X,p)\Big]\Bigg\rbrace,
  \label{eq:TE:p_0_F_LO}
  \\
  \label{eq:TE:p_0_rho_LO}
    \pDiff{p_0} [\tilde \rho(X,p)]
    &=& \Ci\sigma_2\Bigg\lbrace\Big[\frac {\F p^2}{2m}+\Sigma^{(0)}(X)\Big] \tilde \rho(X,p) 
    \nonumber\\
    && + \frac 1 2 \Big[\tilde \Sigma^+(X,p)\tilde \rho(X,p)+\tilde \rho^+(X,p)\tilde\Sigma^\rho(X,p)\Big]\Bigg\rbrace,
  \end{eqnarray} 
as well as NLO equations:
  \begin{eqnarray}
    \pDiff{p_0}[\tilde F] 
    &=& \mathrm{LO} -\frac{\sigma_2}4\Big[
        \lbrace 
          \tilde\Sigma^\rho(X,p),\tilde F(X,p) 
        \rbrace_0
       -\lbrace 
          \tilde \Sigma^F(X,p),\tilde \rho(X,p)
        \rbrace_0 \Big],
  \label{eq:TE:p_0_F_NLO}
  \\
  \label{eq:TE:p_0_rho_NLO}
    \pDiff{p_0} [\tilde \rho] 
    &=&\; \mathrm{LO},
  \end{eqnarray} 
where the operator $\pDiff{p_0}$ is defined in analogy to Eq.~(\ref{eq:TE:limits1}):
  \begin{eqnarray}
    \pDiff{p_0}[\tilde F(X,p)] 
    &=& \int_{-2X_0}^{2X_0} \diff s_0\, \int\diff^3 s \, \e^{\Ci ps} \pdiff{s_0} F(X,s ) 
    \nonumber \\
    &=& -\Ci p_0 \tilde F(X,p)+\Big[\e^{\Ci p_0 s_0} F(X, s_0, \F p)\Big]_{s_0=-2X_0}^{2X_0}.
\label{eq:TE:limits2}
  \end{eqnarray} 
%

\subsection{Noninteracting gas}
\label{app:NonintGas}
\noindent
In order to obtain an estimate for the form of the correlation functions in the $X_0$-$s_0$-plane, we derive analytical solutions for the interaction free case. 
This also recovers the fluctuation dissipation relation between $F$ and $\rho$.

Since the previously made approximations only affect the interaction terms, the leading-order transport equations are equivalent to the 2PI dynamic equations as can be seen by setting $\tilde \Sigma(X,p)\equiv 0$. 
For a homogeneous system, spatial derivatives with respect to the centre coordinates disappear. 
Moreover, the momentum distribution $f(\F p)=\int\diff^3 s\, \e^{-\Ci \F p\cdot\F s}f(\F s)$ is real-valued. 
From equations (\eqref{eq:TE:p_0_F_LO}), (\eqref{eq:TE:p_0_rho_LO}), we obtain
  \begin{eqnarray}
    -\Ci p_0\tilde F(X,p)
    &=& \Ci\sigma_2\frac{\F p^2}{2m}\tilde F(X,p) 
    - \Big[\e^{\Ci p_0 s_0}F(X,s_0,\F p)\Big]_{s_0=-2X_0}^{2X_0}, 
\label{eq:TE:disprel_F}
    \\
    -\Ci p_0\tilde\rho(X,p)
    &=& \Ci\sigma_2\frac{\F p^2}{2m}\tilde\rho(X,p) 
    - \Big[\e^{\Ci p_0 s_0}\rho (X,s_0,\F p) \Big]_{s_0=-2X_0}^{2X_0}.
\label{eq:TE:disprel_rho}
  \end{eqnarray} 
These relations, together with the initial conditions (\eqref{eq:IniCondRho}) imply that
  \begin{equation}
    \rho(X, s_0, \F p) 
    = -\Ci\sigma_2\exp\Big(\Ci\frac{\F p^2}{2m}\sigma_2 s_0\Big), 
\label{eq:TE:ref1}
  \end{equation}
and, in frequency space,
  \begin{eqnarray}
    \tilde\rho(X,p_0, \F p) 
      &=&\,\frac \Ci 2 \Bigg[\delta\left(\frac{\F p^2}{2m}-p_0\right)
                          +\delta\left(\frac{\F p^2}{2m}+p_0\right)\Bigg]
     \nonumber\\
      && ~~~ \times
      \Big[(\theta(p_0)-\theta(-p_0))\, \opUnit + \sigma_2\Big],
\label{eq:TE:analyticalRho}
  \end{eqnarray} 
where $\opUnit$ denotes the $2\times2$ unit matrix.
Since the equations (\ref{eq:TE:disprel_F}) and (\ref{eq:TE:disprel_rho}) for $\tilde F$ and $\tilde\rho$ are identical, one obtains, with the initial momentum distribution $f(\F p)=n(\F p)+1/2$, see Eq.~(\eqref{eq:IniCondF}), the fluctuation dissipation relation (\ref{eq:TE:fluctuationdissipation2}) between $\tilde F$ and $\tilde\rho$, such that
  \begin{equation}
    F(X, s_0, \F p) 
    = f(\F p)\exp\Big(\Ci\frac{\F p^2}{2m}\sigma_2 s_0\Big). 
\label{eq:TE:ref2}
  \end{equation}
%

\section{Self-energies for a homogeneous gas in one spatial dimension}
\label{app:SelfEn1Dhom}
\noindent
In this appendix, we provide the momentum-space self-energies $\Sigma^{F,\rho}_{ab}(x_0,y_0;p)$ in 1+1 dimensions which enter the dynamical equations (\ref{eq:DE:Fp}), (\ref{eq:DE:rhop}).
From Eqs.~(\ref{eq:DE:SigmaF}) to (\ref{eq:DE:IRho}), one obtains, for a homogeneous system, with $\phi_a\equiv0$, by Fourier transformation:
\begin{eqnarray}
\label{eq:SigmabarFpofFrho}
  \Sigma^F_{ab}(x_0,y_0,p)
  &=& -g\int_k\Big[I^F(x_0,y_0,p-k)F_{ab}(x_0,y_0,k)
  \nonumber\\
  &&\qquad
      -\frac{1}{4}I^\rho(x_0,y_0,p-k)\,\rho_{ab}(x_0,y_0,k)\Big],
  \\ 
\label{eq:SigmabarrhopofFrho}
  \Sigma^\rho_{ab}(x_0,y_0,p)
  &=& -g\int_k\Big[I^\rho(x_0,y_0,p-k)F_{ab}(x_0,y_0,k)
  \nonumber\\
  &&\qquad
                 +I^F(x_0,y_0,p-k)\rho_{ab}(x_0,y_0,k)\Big],
\end{eqnarray} 
where
\begin{eqnarray}
\label{eq:IFp}
  &&I^F(x_0,y_0,p)
  =\frac{g}{2}\int_{k}\Big\{
    F_{ab}(x_0,y_0,p-k)F_{ab}(x_0,y_0,k)
  \nonumber\\
  &&\qquad\qquad\qquad\qquad
    -\frac{1}{4}\rho_{ab}(x_0,y_0,p-k)\rho_{ab}(x_0,y_0,k)
  \nonumber\\
  &&\qquad
    -\int_{k'}\Big[\int_0^{x_0}dz_0\,I^\rho(x_0,z_0,p-k)\,
              \Big(F_{ab}(z_0,y_0,k-k')F_{ab}(z_0,y_0,k')
  \nonumber\\
  &&\qquad\qquad\qquad
    -\frac{1}{4}\rho_{ab}(z_0,y_0,k-k')\rho_{ab}(z_0,y_0,k')\Big)
  \nonumber\\
  &&\qquad
    +2\int_0^{y_0}dz_0\,I^F(x_0,z_0,p-k)\,
	         F_{ab}(x_0,y_0,k-k')\rho_{ab}(x_0,y_0,k')\Big]\Big\},   
\end{eqnarray} 
\begin{eqnarray}
\label{eq:Irhop}
  &&I^\rho(x_0,y_0,p)
  = g\int_{k}\Big\{
    F_{ab}(x_0,y_0,p-k)\rho_{ab}(x_0,y_0,k)
  \nonumber\\
  &&\qquad
    -\int_{k'}\int_{y_0}^{x_0}dz_0\,I^\rho(x_0,z_0,p-k)\,
                 F_{ab}(z_0,y_0,k-k')\rho_{ab}(z_0,y_0,k')\Big\},
\end{eqnarray} 
Here, $\int_k\equiv(2\pi)^{-1}\int dk$ denotes the one-dimensional momentum integral.

\end{appendix}
\ \\

\section*{References}
\bibliographystyle{bibtex/JHEP-2.bst}
\bibliography{bibtex/mybib,bibtex/additions}

\end{document}